\documentclass[multphys,vecphys]{svmult}

\usepackage{makeidx}         
\usepackage{graphicx}        
\usepackage{multicol}        
\usepackage[bottom]{footmisc}

\makeindex             

\def\beq{\begin{equation}}
\def\eeq{\end{equation}}

\def\undersim#1{\mathop{\vtop{\ialign{##\crcr
$\hfil\displaystyle{#1}\hfil$\crcr\noalign
{\kern0pt\nointerlineskip}\hbox{$\hfil\sim\hfil$}\crcr
}}}}

\def\msol{\mathrm{\ M}_\odot}

\begin{document}

\title*{Simulations of globular clusters merging in galactic nuclear regions}
\author{P. Miocchi\inst{1,2} \and R. Capuzzo Dolcetta\inst{2} \and
P. Di Matteo\inst{3,2}}
\institute{INAF - Osserv. di Teramo, Via M. Maggini, 64100 -- Teramo, Italy
\texttt{miocchi@uniroma1.it}
\and
Dipartimento di Fisica, Universit\'a di Roma ``La Sapienza",
P.le Aldo Moro, 2, I00185 -- Rome, Italy\\
\texttt{roberto.capuzzodolcetta@uniroma1.it}
\and
LERMA  - Observ. de Paris, 61, Av. de L'Observatoire,
75014 -- Paris, France
\texttt{paola.dimatteo@obspm.fr}}
%
%
\maketitle
\abstract
We present the results of detailed $N$-body simulations regarding the interaction of
four massive globular clusters in the central region of a
triaxial galaxy. The systems undergo a full merging event, producing a sort of
`Super Star Cluster' (SSC) whose features are close to those of a
superposition of the individual initial mergers. In contrast with other similar
simulations, the resulting SSC structural parameters are located along
the observed scaling relations of globular clusters. These findings seem to
support the idea that a massive
SSC may have formed in early phases of the mother galaxy
evolution and contributed to the growth of a massive nucleus.

\section{Introduction}
\label{intro}
In the paper \cite{miocchi05}, we analyzed the head-on
collision between two globular clusters (GCs) moving on quasi-radial orbits
in the galactic central
region, with the aim, also, to understand how effective is the
cluster tidal destruction.
One of the main findings was that
sufficiently compact clusters (initial King
concentration parameter $c\geq 1.6$) keeps bound a substantial amount of their
mass up to the complete orbital decaying. Another
important result is that the orbital energy dissipation due to the tidal
interaction is of the same order of that caused by dynamical friction (df).
In light of these results, and given that df was shown to
be important indeed in segregating massive GCs in triaxial potentials
(\cite{tremetal75, capdol93}), a natural further step in our
investigation program is to study the possible merging of a set of GCs
decayed in the central
galactic region and the relevant characteristics
of the super star cluster (SSC) they eventually form.

Furthermore, this topic is connected with the problem
of the origin and formation history of the various kinds of SSCs
recently observed in form of ultracompact dwarf galaxies
(\cite{drink00}), young massive clusters in starburst regions of
interacting galaxies (\cite{mara04}), and nuclear
clusters in the central regions of early-type spirals (\cite{Bok04,walch05}).
One of the most debated questions is whether they formed
from the condensation of primordial gas clouds or through the
``dissipationless'' merging of smaller subunits.
Recent indications coming from some $N$-body simulations
(\cite{fell02,baumg03,bekki04})
are in favour of this latter hypothesis.
In this work we studied the
merging process among massive GCs already decayed in the
central region of a triaxial galaxy, including tidal
distortion and df.


\section{The model}
\label{model}
The galactic environment in which GCs are embedded is represented by
an analytical potential derived from the self-consistent \textit{triaxial}
model described in \cite{zeeuw} (see also \cite{miocchi05}).
The galactic core radius is $200$ pc, the core
mass is $3\times 10^9\msol$ with a central density $\rho_{g0}=375\msol$pc$^{-3}$.
The deceleration due to df on the stellar motion is considered through
a generalization of the Chandrasekhar formula
to the triaxial case (\cite{PCV}).
The four simulated GCs have an initial internal distribution sampled
from King isotropic models with total mass,
central velocity dispersion and half-mass radius ranging, rispectively, in
$M\simeq 42$--$54\times 10^6\msol$, $\sigma\simeq 25$--$36$ km s$^{-1}$,
$r_c\simeq 11$--$19$ pc.
The reference frame has the origin at the
galactic center and the $x$ and $z$ axes are, respectively, along the
maximum and minimum axis of the triaxial ellipsoid.
The clusters were initially located within $100$ pc
from the galactic center.
We represented each GC with $N=2.5\times 10^5$ `particles', simulating
their dynamics by means of the parallel `ATD' treecode
(\cite{bib2,miocchi05}).

\section{Results}
\label{results}

\begin{figure}
\centering
\includegraphics[height=11.5cm]{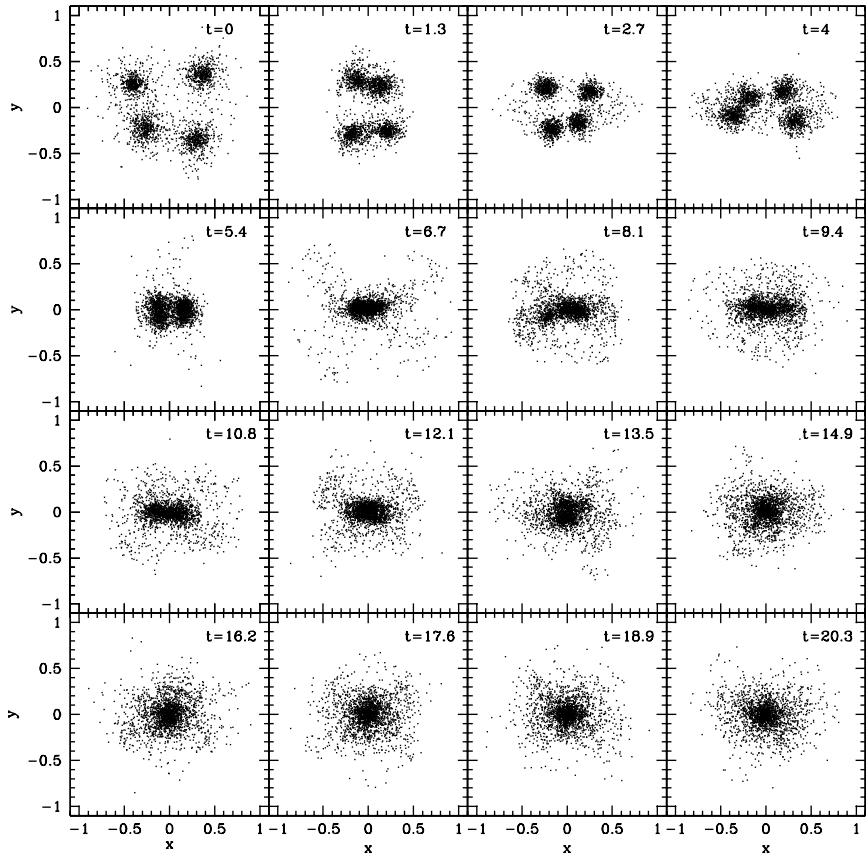}
%
%
\caption{Snapshots of the merging event (projection on the $x$--$y$ plane). The labelled time is in unit of $0.8$ Myr. 
Each snapshot size is $\sim 400$ pc.}
\label{snap1}       
\end{figure}

From Fig.~\ref{snap1} it can be clearly seen that the merging occurs rather quickly,
i.e. in less than 20 galactic core crossing time ($\sim 14$ Myr),
after which the resulting system attains soon a dynamical equilibrium configuration,
as confirmed by the evolution of the lagrangian radii that are
nearly constant throughout the duration of the simulation ($\sim 35$ Myr)

\begin{figure}
\centering
\includegraphics[height=8.5cm]{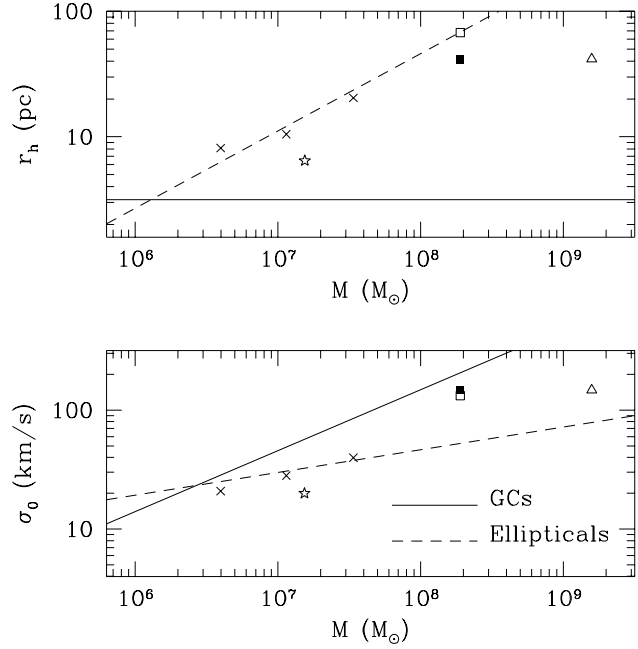}
%
%
\caption{Scaling relations for the last configuration of our SSC.
The best fit relations for GCs
and elliptical galaxies are also reported. The squares indicate the location of
our SSC in the case where df is present (filled) or not (open), and when
$M$ is estimated by the virial relation (open triangles).
The merging remnants locations
for some of the cases simulated by \cite{bekki04} (crosses) and by
\cite{fell02} (stars) are also plotted.}
\label{sigma_m}       
\end{figure}

The final SSC morphology is that of an axisymmetric ellipsoid (axial ratios
$1.4$ : $1.4$ : $1$, ellipticity $\simeq 0.3$) without figure rotation.
Nearly all the progenitors mass ends up into the SSC, which shows
a central velocity dispersion $\sigma_0\simeq 150$ km s$^{-1}$ and a
half-mass radius of $r_h\simeq 40$ pc. This means that the SSC is located
much closer to the $\sigma_0$--$M$ scaling relation followed by GCs
than to that of elliptical galaxies (Fig.~\ref{sigma_m} lower panel),
contrarily to what found by \cite{bekki04,fell02}.
This is probably due to that our system is located deep inside
the galaxy potential well.
Indeed, if one uses the virial relation to estimate
the system total mass, $M=7.5 \sigma_0^2r_h/G$ (\cite{kiss}), then
the location approaches the elliptical galaxies scaling law (Fig.~\ref{sigma_m}).

\begin{figure}
\centering
\includegraphics[height=6cm]{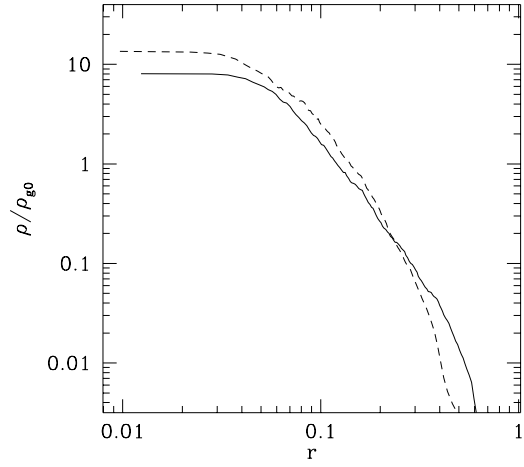}
%
%
\caption{Radial density profile of the SSC (solid line), compared with the profile
given by the sum of the initial density distributions of the progenitors clusters
(dashed line). The distance $r$ (in units of $200$ pc) is to the galactic center.}
\label{dens}       
\end{figure}

The radial density profile of the SSC (Fig.~\ref{dens}) is similar to
that given by the `sum' of the initial profiles of the progenitors,
with a central density a factor only $1.7$ smaller and a slightly shallower
profile outwards.
This is in the direction of supporting the validity of the nuclear accretion
model, suggested by ~\cite{capdol93}.
Finally, we found that the SSC formed in the simulation without df,
reaches an equilibrium state with similar morphological features, though in a
nearly doubled time-scale and with a $\sim 3$ times lower central density.


%
%
%
%
%
%
%
%
%
%

%
%

%
%
%




\printindex
\end{document}